\documentclass[twocolumn,prb,superscriptaddress,showpacs,floatfix]{revtex4}
\usepackage{amsfonts}
\usepackage{amssymb}
\usepackage{amsmath}
\usepackage{graphicx}
\usepackage{dcolumn}
\usepackage{bm}
\usepackage{flafter}
\usepackage[normalem]{ulem}

\begin{document}

\title{Anisotropic Dirac electronic structures of {\it A}MnBi$_2$ ({\it A}=Sr, Ca)}

\author{Geunsik Lee }
\email{maxgeun@postech.ac.kr}
\affiliation{Department of Chemistry, Pohang University of Science and Technology, San 31, Hyojadong, Namgu, Pohang 790-784, Republic of Korea}
\author{Muhammad A. Farhan}
\affiliation{Department of Chemistry, Pohang University of Science and Technology, San 31, Hyojadong, Namgu, Pohang 790-784, Republic of Korea}
\author{Jun Sung Kim}
\affiliation{Department of Physics, Pohang University of Science and Technology, San 31, Hyojadong, Namgu, Pohang 790-784, Republic of Korea}
\author{Ji Hoon Shim}
\email{jhshim@postech.ac.kr}
\affiliation{Department of Chemistry, Pohang University of Science and Technology, San 31, Hyojadong, Namgu, Pohang 790-784, Republic of Korea}
\affiliation{Department of Physics, Pohang University of Science and Technology, San 31, Hyojadong, Namgu, Pohang 790-784, Republic of Korea}
\affiliation{Devision of Advanced Nuclear Engineering, Pohang University of Science and Technology, San 31, Hyojadong, Namgu, Pohang 790-784, Republic of Korea}
\date{\today}

\begin{abstract}
Low energy electronic structures in {\it A}MnBi$_2$ ({\it A}=alkaline earths) are
investigated using a first-principles calculation and a tight binding method.
An anisotropic Dirac dispersion is induced by the checkerboard arrangement of {\it A} atoms 
above and below the Bi square net in {\it A}MnBi$_2$.
SrMnBi$_2$ and CaMnBi$_2$ have a different kind of Dirac dispersion 
due to the different stacking of nearby {\it A} layers, where 
each Sr (Ca) of one side appears at the overlapped (alternate) position of the same element at the other side.
Using the tight binding analysis,
we reveal the chirality of the anisotropic Dirac electrons as well as the sizable spin--orbit
coupling effect in the Bi square net. 
We suggest that the Bi square net provides a platform for the interplay between
anisotropic Dirac electrons and the neighboring environment such as magnetism and structural changes.
\end{abstract}


\maketitle

\section{Introduction}
Low energy electrons obeying the relativistic Dirac equation have been reported in 
so-called Dirac materials such as graphene.\cite{natmat-geim,prl-theo}
Many intriguing physical properties of Dirac fermions have been investigated both theoretically
and experimentally, including the unconventional
quantum Hall effect, Klein tunneling, suppressed back scattering, and so on.\cite{rev-graph} 
Dirac materials are also reported in
topological insulators,\cite{TI,TI1,TI2}
iron pnictides,\cite{FeSC,FeSC1} organic conductors,\cite{orgcon} inverse perovskite,\cite{invper}
and the VO$_2$--TiO$_2$ interface,\cite{Pick2009} with the list is rapidly growing these days.
The effective Hamiltonian in all those systems is characterized by the Pauli matrices,
which is essential to manifest spinor-related phenomena.

Recently, an anisotropic Dirac cone in SrMnBi$_2$ was found by a density functional theory (DFT) calculation
and confirmed by angle revolved photoemission spectroscopy and quantum oscillations.\cite{Park2012}
It is characterized by linear dispersions with strong anisotropy in the momentum-dependent Fermi velocity.
Such a Dirac band near the Fermi level mainly results from the Bi square net layer,\cite{Moro2011} 
and is responsible for a finite Berry phase in the Shubnikov--de Haas oscillations\cite{Park2012} and the  
linear-field dependence of the magnetoresistance.\cite{Petro-Sr}
Similar behavior was also found in CaMnBi$_2$ with the same Bi square net.\cite{Petro-Ca,Chen2012}  
This indicates that the low energy electron motion in the Bi square {net can be} described by the Dirac Hamiltonian. 
So far, however, the origin of the Dirac fermions in the Bi squre {net has}
not yet been investigated clearly.
So, this paper is devoted to studying the origin of the anisotropic Dirac electrons observed in SrMnBi$_2$ and CaMnBi$_2$.

This paper is organized as follows.
Section II explains the crystal structure of the two compounds 
and the details of the DFT calculation. 
In Section III, the DFT band structures containing the Dirac dispersion are presented.
In Section IV, we clarify the mechanism of the anisotropic Dirac dispersion as well as the chirality
of the Dirac electron by using a tight binding (TB) method. 
Also, the effect of the spin--orbit coupling (SOC) {of heavy Bi atoms} will be {investigated}.
Finally, we conclude our paper in Section V.

\begin{figure}[b]
\begin{center}
\includegraphics[width=8.5cm,angle=0]{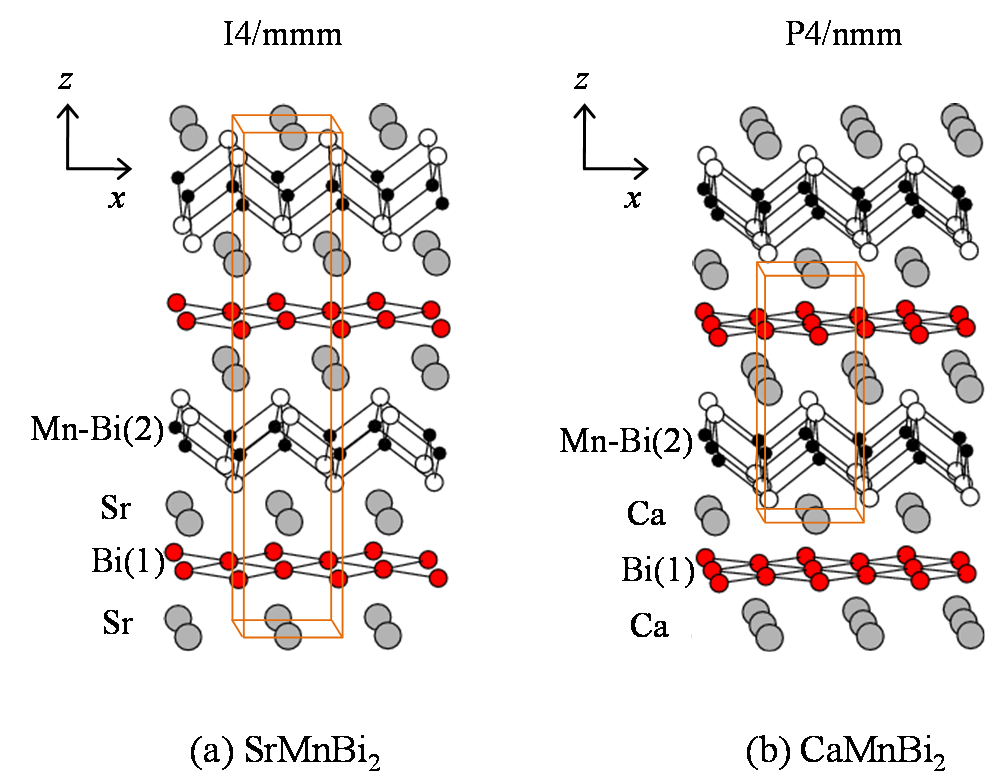}
\caption{\label{DOS} (color online) 
Crystal structures of (a) SrMnBi$_2$ and (b) CaMnBi$_2$,
whose crystal symmetries and atomic positions are listed in Table I.
Bi(1) and Bi(2) indicate the first and second types of Bi atoms, respectively.
The orange (bright gray) line indicates the conventional unit cell.
The primitive unit cell, which is used for the DFT calculation, has the same volume as that of the conventional unit cell for (b) but for (a), it is one-half of that.}
\end{center}
\end{figure}

\section{Calculation method}
Figure 1 shows the crystal structures of SrMnBi$_2$ (SG 139, I4/mmm)
and CaMnBi$_2$ (SG 129, P4/nmm).
Both compounds contain a square net of Bi atoms
which are indicated by the red (dark gray) balls. 
{The Sr or Ca atoms are located at the pyramid top of four base Bi atoms,}
where the vertical distance varies from 2.5 to 2.7 {\AA}.
{Each Sr or Ca layer has a checkerboard ordering with respect to the Bi square net.}
There exist other buffer layers containing Mn-Bi tetrahedrons.
So there are two types of Bi atoms in the unit cell:
Bi(1) in the square net and Bi(2) in the Mn-Bi layer.

The stacking configuration of the two alkaline earth atomic layers above and below the Bi square net
is different for SrMnBi$_2$ from that of CaMnBi$_2$.
As one can see in Fig. 1 or more schematically in Fig. 5, 
each Sr (Ca) of one side appears at the overlapped (alternate) position of the same element at the other side.
Due to this difference, SrMnBi$_2$ (CaMnBi$_2$) has a body-centered (primitive) tetragonal 
Bravais lattice. 
The experimental structural parameters of CaMnBi$_2$ and SrMnBi$_2$ are listed in Table I.

\begin{table}[t]
\caption{\label{table1} 
Experimental structural parameters used for {the DFT calculation}. 
Bi(1) {and} Bi(2) indicate the first {and} second types of Bi atoms, {respectively}, as shown in Fig. 1.}
\begin{tabular}{p{2.0cm}p{3.2cm}p{3cm}}
\hline
compound & CaMnBi$_2$ & 
           SrMnBi$_2$ \\
\hline
space group & P4/nmm  (129) & I4/mmm (139) \\
{\it a} ({\AA}) & 4.50 & 4.58 \\
{\it c} ({\AA}) & 11.08 & 23.13 \\
Ca, Sr         & 2c (0.25, 0.25, 0.724) &  4e (0.0, 0.0, 0.1143) \\
Mn             & 2a (0.75, 0.25, 0.0) &    4d (0.0, 0.5, 0.25)   \\
Bi(1)          & 2b (0.75, 0.25, 0.5) &    4c (0.0, 0.5, 0.0)   \\
Bi(2)          & 2c (0.25, 0.25, 0.1615) & 4e (0.0, 0.0, 0.3265) \\
\hline
references & Ref. \onlinecite{camnbi2} & Ref. \onlinecite{srmnbi2} \\
\hline
\end{tabular}
\end{table}

The band structure calculations are performed
with the full-potential linearized augmented plane-wave method implemented in the WIEK2K package.\cite{wien2k}
The generalized gradient approximation (GGA) by Perdew--Burke--Ernzerhof (PBE) is used for the exchange-correlation potential.\cite{pbe}
The radius of the muffin-tin is set to 2.5 a.u. for all atoms.
For the charge self-consistent calculation, the number of $\mathbf{k}$ points used in the full Brillouin zone is 1000.
Due to the magnetic Mn ions, the spin polarized calculation is carried out.
Also the SOC is considered due to the presence of heavy Bi atom.

\section{DFT results}

\begin{table}[b]
\caption{\label{table1} 
Calculated total energies of the ferromagnetic (FM) and the stripe-type antiferromagnetic (sAFM)
configurations with respect to the 
checkerboard-type antiferromagnetic (cAFM) configuration. The unit is eV per formula unit.}
\begin{tabular}{p{2.0cm}p{2.cm}p{2cm}p{2cm}}
\hline
   & cAFM & sAFM & FM \\
\hline
SrMnBi$_2$ & 0.0 & 0.09 & 0.55 \\
CaMnBi$_2$ & 0.0 & 0.08 & 0.29 \\
\hline
\end{tabular}
\end{table}

\begin{figure}[t]
\begin{center}
\includegraphics[width=8.5cm]{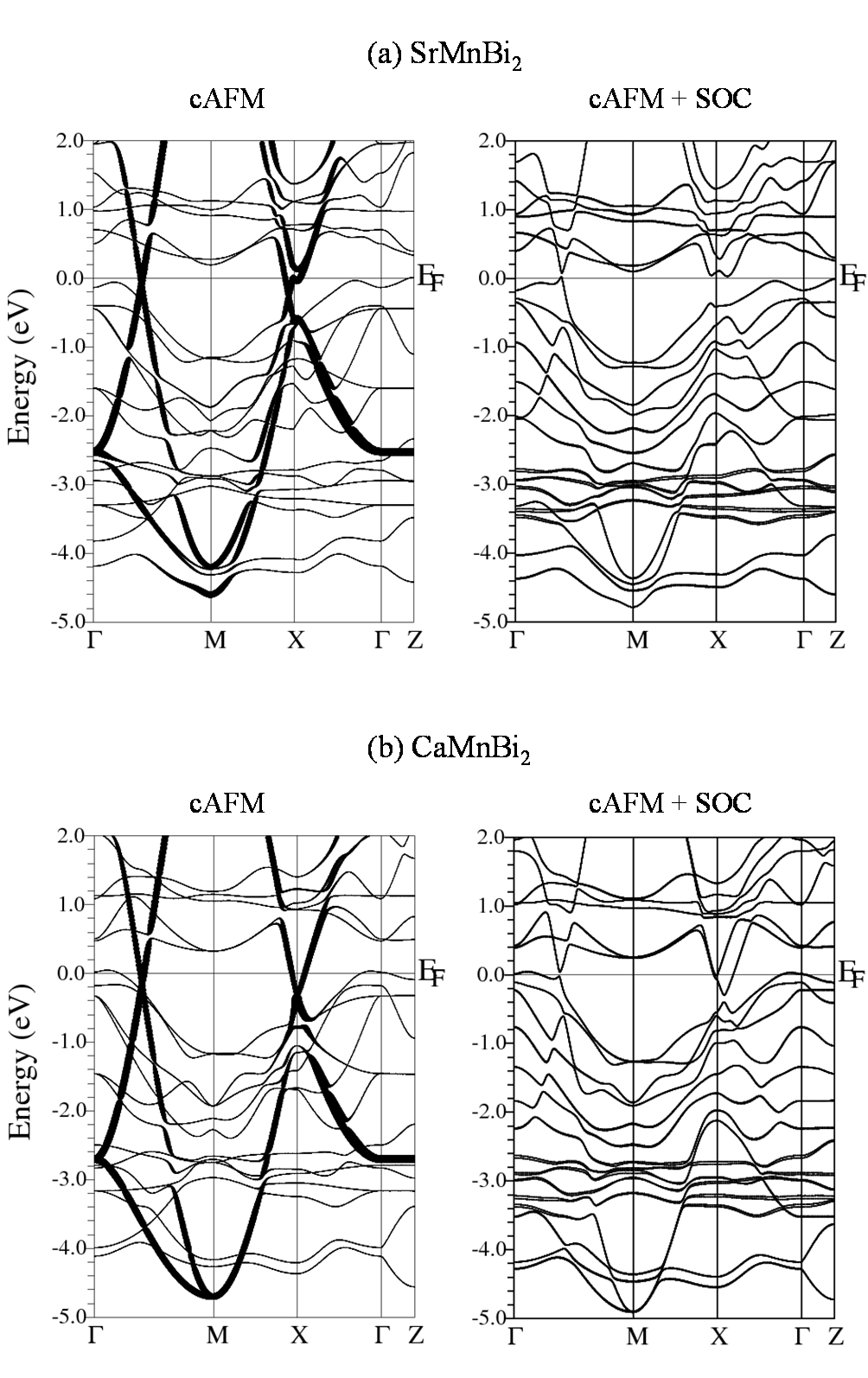}
\caption{\label{band} 
Band structures of (a) SrMnBi$_2$, (b) CaMnBi$_2$.
The checkerboard-type antiferromagnetic (cAFM) order is assumed, and 
cAFM+SOC means the inclusion of the spin--orbit coupling (SOC). 
The contribution from the Bi $p_x$ and $p_y$ orbitals is indicated by the size of the points in the cAFM results.}
\end{center}
\end{figure}

In $A$MnBi$_2$ ($A$=Sr and Ca), Mn$^{2+}$ has a $3d^5$ electron configuration.
The calculated spin magnetic moment within the muffin-tin sphere is about 4 $\mu_B$ for both compounds,
which shows negligible changes under different magnetic configurations of Mn spins.
In order to find the magnetic ground state, the total energies are calculated for
{various magnetic configurations such as} ferromagnetic (FM), checkerboard-type antiferromagnetic (cAFM),
and the stripe-type antiferromagnetic (sAFM) configurations.
Their relative energies are listed in Table II.
The ground state is found to be cAFM for both CaMnBi$_2$ and SrMnBi$_2$.
The relative stability listed in Table II is qualitatively consistent with the literature.\cite{Moro2011}
The associated cAFM transition will be related to {the magnetic transition} observed at {$\sim$} 290 K in the experiments.\cite{Park2012} 
Under the cAFM order, the interlayer exchange interaction is so weak due to the large MnBi interlayer
distance, and its contribution is as small as the order of magnitude of the energy tolerance ($\sim$ 1 meV/f.u.).
In our calculation {we assumed an antiferromagnetic interlayer ordering of the Mn spins} along the $z$ axis, 
which does not alter the main feature of the band structures.
The  total energy difference in Table II remains almost the same when taking into consideration the SOC.

As explained, two Sr (or Mn-Bi(2)) layers adjacent to the Bi(1) square net in SrMnBi$_2$ 
overlap when 
projected along the $z$ axis, while the adjacent Ca layers are alternate in CaMnBi$_2$.
The total energy calculation is consistent with the experimental observation that SrMnBi$_2$ favors the 
overlapping type, while CaMnBi$_2$ favors the alternate type.
So it is likely that the stacking type is mainly determined by the ionic size.
Interestingly, another structurally related compound SrMnSb$_2$ with the same stacking type as CaMnBi$_2$ is 
known to exhibit the chain-type reconstruction of the Sb square net.\cite{srmnsb2}
However, CaMnBi$_2$ does not show any distortion in the Bi square net. 
The absence of such distortion in the Bi square net in CaMnBi$_2$ is mainly due to the SOC.
In fact we found that soft phonon modes of 42 cm$^{-1}$ (5.3 meV) lead to a chain-type
reconstruction in CaMnBi$_2$ without including the SOC, but they disappear when the SOC is included. 
Thus, strong SOC has an important role in the suppression of the chain-type distortion in CaMnBi$_2$.

Figure 2 shows the calculated band structures of SrMnBi$_2$ and CaMnBi$_2$ both without and with the SOC.
Both compounds show spin-polarized Mn-driven bands near $-3.0$ eV and 1.0 eV with majority and minority spins, respectively.
As indicated by the size of the points, the bands near the Fermi level arise mostly from the Bi(1) $p$ orbitals
which are weakly hybridized with Sr or Ca $d$ orbitals.
Near the Fermi level, two linear bands cross 
at a certain wavevector, $\mathbf{k_0}$, along $\Gamma$ to M.
When the SOC is included, 
the two linearly crossing bands change to quasi-linear bands with a small SOC-induced 
gap of 0.05 eV. The SOC-induced gap is much larger at the X point, so most of the states near the X point are removed from the Fermi level.
This is because 
the electronic states have a smaller energy difference near the X point than $\mathbf{k_0}$, 
and thus the SOC splitting is more sensitive to the nonvanishing SOC interaction.
This will be shown more clearly in Section IV.C, based on the TB analysis and perturbation theory.

\begin{figure}[t]
\begin{center}
\includegraphics[width=8.5cm]{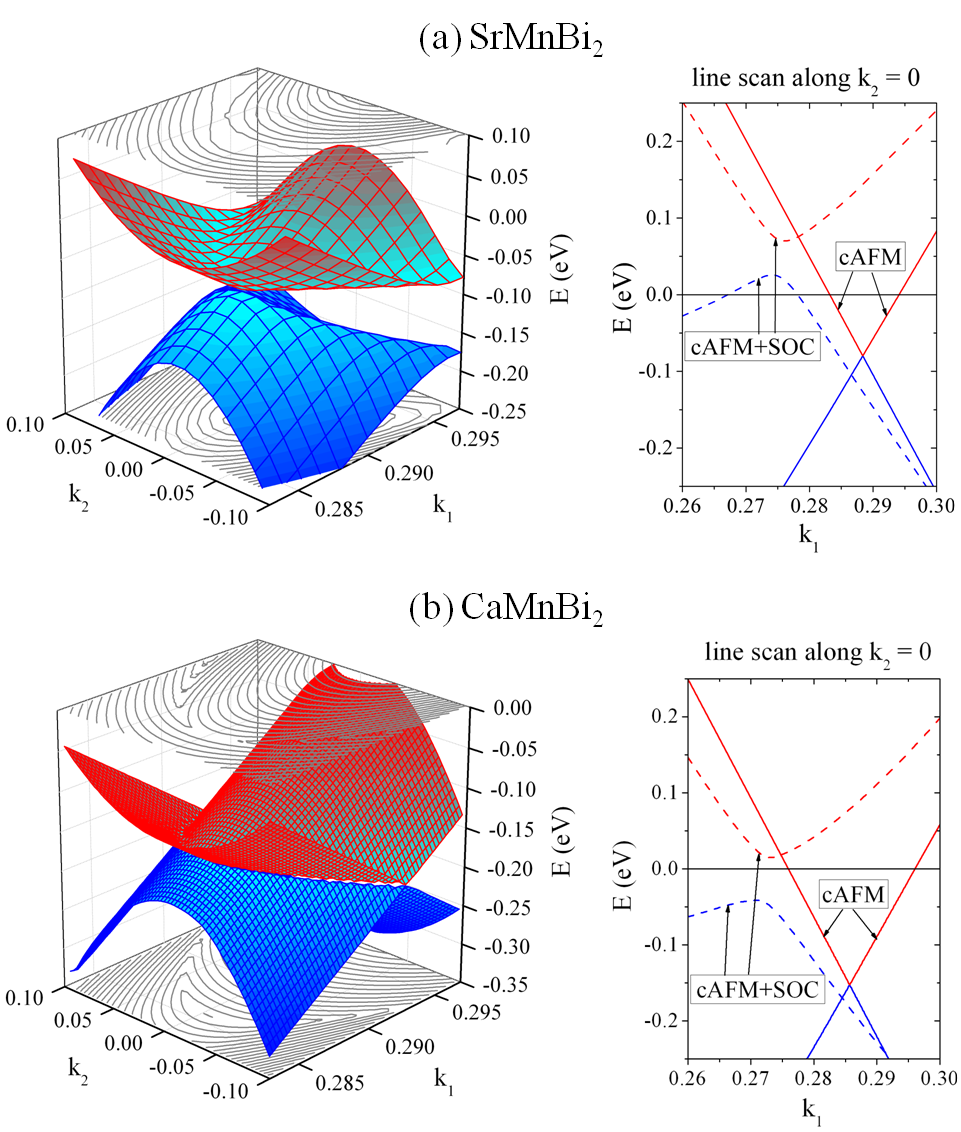}
\caption{\label{band} (color online) 
Band energy surface (left hand side) and dispersion curve (right hand side) 
near the Dirac point for (a) SrMnBi$_2$ and (b) CaMnBi$_2$.
$k_1$ is a coordinate along the $\Gamma$-M direction, and its perpendicular axis is $k_2$, 
defined by $k_1=(k_x+k_y)/\sqrt{2}$ and $k_2=(-k_x+k_y)/\sqrt{2}$.
{\it Note the displayed range of $k_1$ and $k_2$: 
$0.283\le k_1 \le 0.297$ and
$-0.10\le k_2 \le 0.10$ in units of $2\pi/a$ with a ratio of $\Delta k_2 /\Delta k_1 \sim 14$.}
The cAFM configuration without SOC is assumed, and the cAFM+SOC result is also shown by the dashed line
in the right hand side.}
\end{center}
\end{figure}

The linearly crossing bands in SrMnBi$_2$ and CaMnBi$_2$ have been ascribed to the observed Dirac fermion-like dispersion.
\cite{Petro-Ca,Chen2012,Moro2011,Park2012,Petro-Sr,Pero-magtherm}
In Fig. 3, we show the energy dispersion near the Dirac point at $\mathbf{k_0}$ without including the SOC. Here $k_1$ and $k_2$ are defined by $k_1=(k_x+k_y)/\sqrt{2}$
and $k_2=(-k_x+k_y)/\sqrt{2}$, where $(k_x,k_y)$ corresponds to $(0,0)$ and $(\pi/a,\pi/a)$ at the $\Gamma$ and M points, respectively.
As shown in Fig. 3(a), the anisotropy, i.e., the ratio of Fermi velocities along the $k_1$ and $k_2$ directions, 
is as high as 50. 
Also the hole and electron bands touch at the Dirac point.
In the case of CaMnBi$_2$, shown in Fig. 3(b), the overall features in the energy dispersion near $\mathbf{k_0}$ are similar to those of SrMnBi$_2$.
However the gap introduced by the hybridization with the states from
the alkaline earth atoms is rather different in CaMnBi$_2$.
For SrMnBi$_2$, the zero-energy gap is found only at the $\mathbf{k_0}$ point between $\Gamma$ and M,
while it is found along a continuous line in the momentum space 
for CaMnBi$_2$. These reflect the important role of the arrangement of the alkaline earth atoms
with respect to the Bi square net, which will be discussed in detail in Section IV.A.
In addition, when the SOC is included, a small gap is introduced at the Dirac point as mentioned above.
However, the essential feature of the anisotropic Dirac dispersion is maintained.

\begin{figure}[t]
\begin{center}
\includegraphics[width=8.5cm]{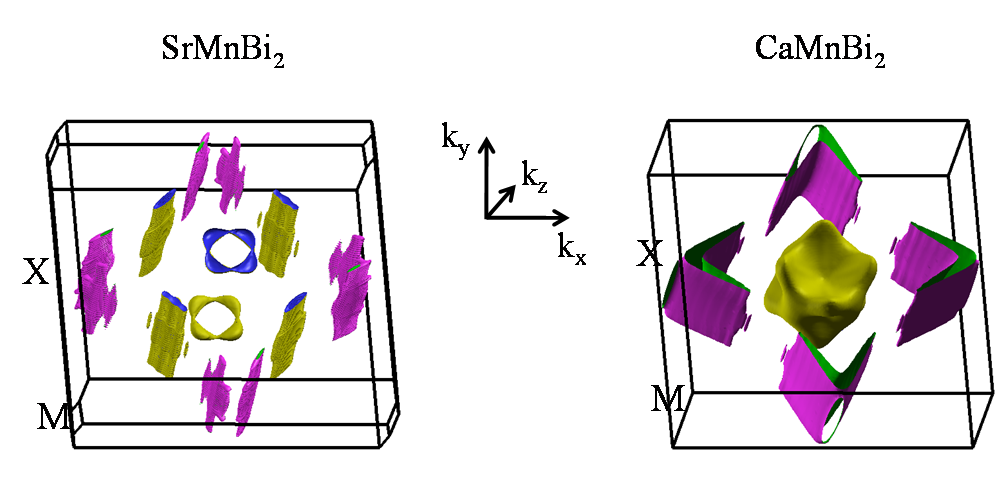}
\caption{\label{FS} (color online) 
Calculated Fermi surfaces of SrMnBi$_2$ and CaMnBi$_2$. 
The direction of the reciprocal axis is provided between them.
The zone center is regarded as the $\Gamma$ point.
The color code indicates the carrier type of the Fermi surfaces.
Yellow and pink surfaces correspond to hole and electron carrier types, respectively, while the opposite face of each Fermi surface is encoded by another color.}
\end{center}
\end{figure}

Figure 4 shows the calculated Fermi surfaces by considering the SOC.
Near $\Gamma$, hole pockets exist in both compounds.
Around X or $\mathbf{k_0}$, there are strongly anisotropic pockects caused by the Dirac bands 
for SrMnBi$_2$ and CaMnBi$_2$.
Also, these are highly two-dimensional, which is consistent with the result of Fig. 2 that 
the Dirac band is mainly caused by Bi $p_x$ and $p_y$ orbitals.
Such Dirac fermions are supposed to dominate the transport property mainly because of their high Fermi velocities.
From our calculation, the carrier type near the X point is the electron type, and near the $\mathbf{k_0}$ point, the hole type.
Since the hole pocket at $\mathbf{k_0}$ in SrMnBi$_2$ is absent in CaMnBi$_2$, 
the hole carrier density is greater in SrMnBi$_2$ than CaMnBi$_2$.
This may be related to
an experimental result that positive (negative) thermopower is observed 
in SrMnBi$_2$ (CaMnBi$_2$).\cite{Pero-magtherm}

\section{TB analysis}
Our {\it ab initio} band structures for SrMnBi$_2$ and CaMnBi$_2$ indicate that
the Dirac-like dispersion is mainly caused by the Bi(1) $p_x$ and $p_y$ orbitals.
Depending on the stacking of Sr or Ca, they show a Dirac point or a continuous band crossing line.
In order to understand the main mechanism for the different band structures,
we carried out a TB analysis on both CaMnBi$_2$ and SrMnBi$_2$.
We construct the TB Hamiltonian of the Bi square net with and without including the interaction with the Sr or Ca atom,
and discuss the mechanism for the Dirac-like electronic structures.
Also the chiral nature of the Dirac electron as well as the effect of the SOC are investigated.

\subsection{Bi square net with $\sqrt{2}\times\sqrt{2}$ unit cell}

\begin{figure}[t]
\begin{center}
\includegraphics[width=8.5cm]{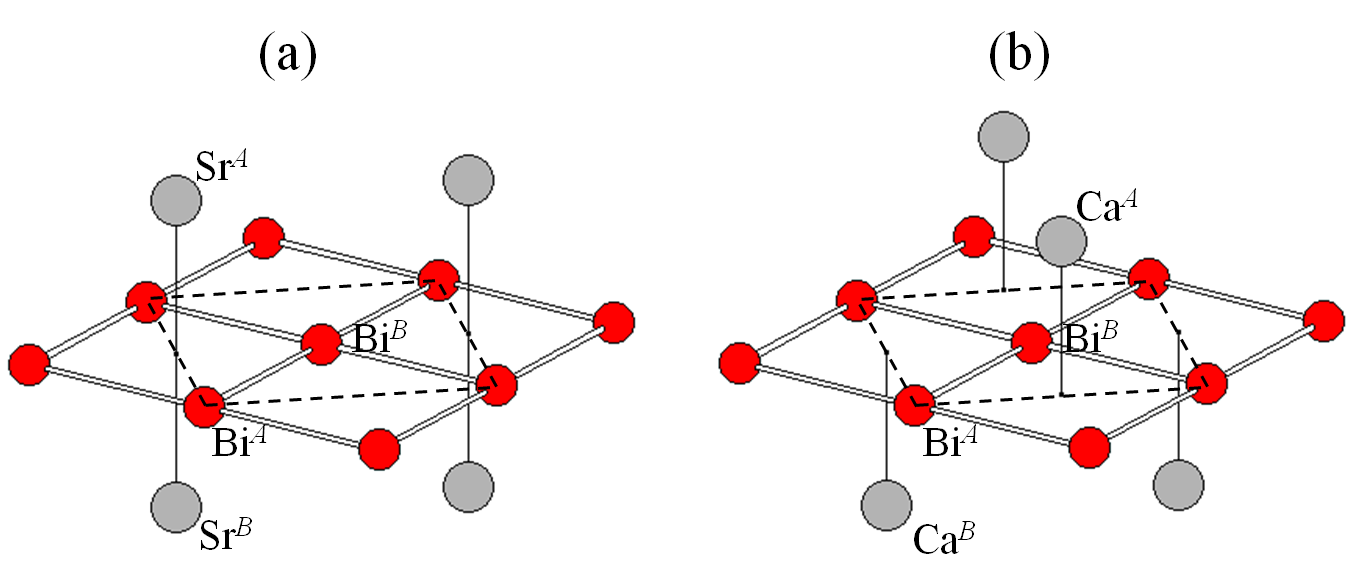}
\caption{ (color online) Bi square net containing (a) Sr and (b) Ca.
The dashed line indicates the $\sqrt{2}\times\sqrt{2}$ unit cell due to the arrangement of the Sr or Ca atoms.
The red and dark gray (bright gray) balls mean Bi and Sr or Ca atoms, respectively.
The atomic coordinates are 
Bi$^A (0,0,0)$, Bi$^B (a/2,a/2,0)$, 
Sr$^A (0,a/2,a/2)$, Sr$^B (0,a/2,-a/2)$, 
Ca$^A (a/2,0,a/2)$, Ca$^B (0,a/2,-a/2)$. }
\end{center}
\end{figure}

\begin{figure}[t]
\begin{center}
\includegraphics[width=8.5cm]{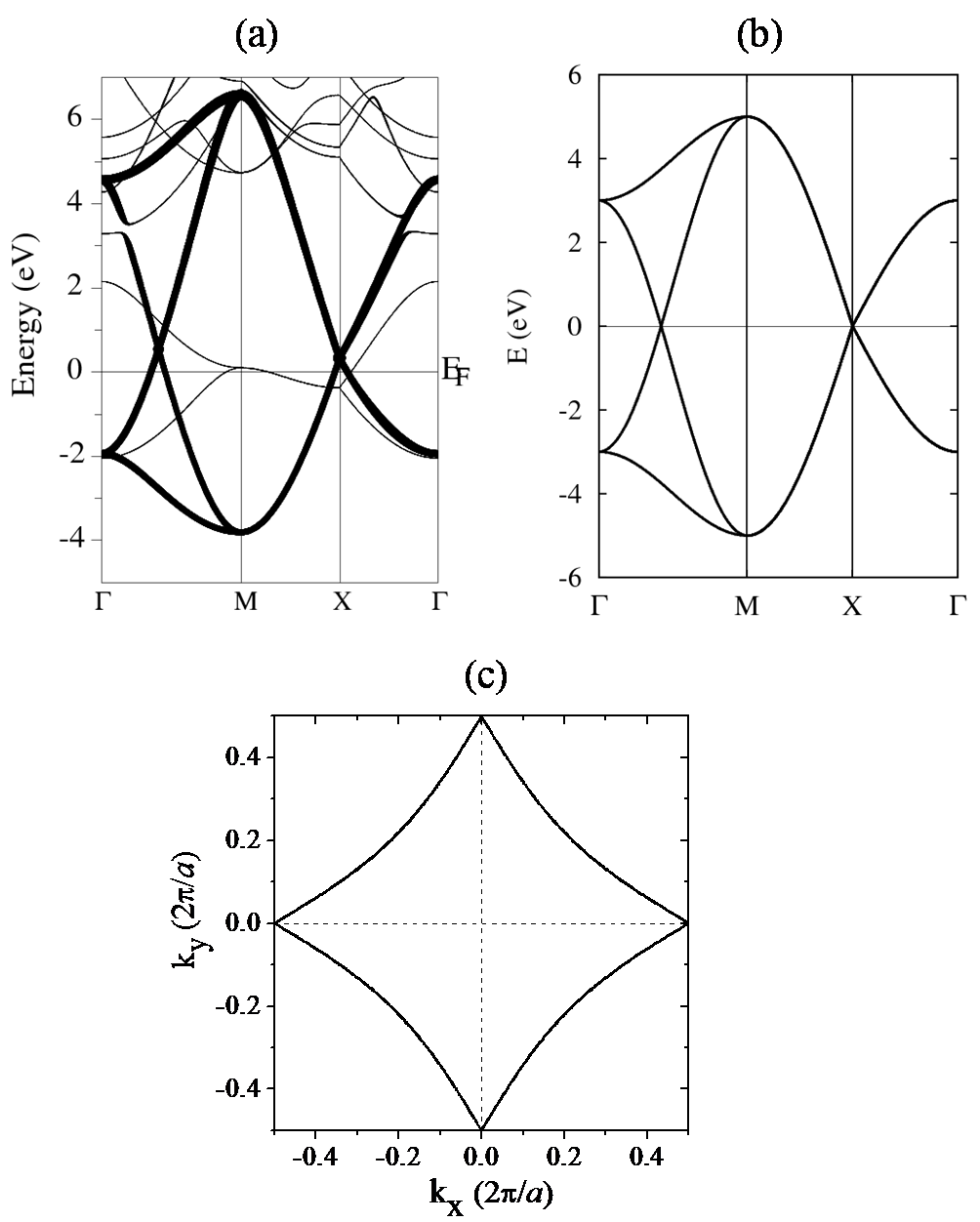}
\caption{The band structures of a single layer Bi square net.
(a) The band structure obtained by the DFT method.
The contribution from Bi 6$p_x$ and 6$p_y$ atomic orbitals are indicated by the size of the points.
(b) The TB band structure considering only $p_x$ and $p_y$ orbitals with the hopping parameters
 $t_{1\sigma}$=2.0 eV, $t_{1\pi}=-0.5$ eV.
(c) Fermi surface plot by the TB method.  }
\end{center}
\end{figure}

Figure 5 represents the Bi square lattice interacting with the Sr or Ca atoms.
Because of the unit cell doubling due to the Sr or Ca atoms, there are two Bi atoms at (0,0,0) and ($a/2$,$a/2$,0)
in the primitive unit cell with the lattice constant $a$.
They are denoted by Bi$^A$ and Bi$^B$, respectively.
First, we study the pristine Bi square net without considering the Sr or Ca atoms.
From the DFT result in the absence of the SOC,
we have checked that the linearly crossing bands are mainly dominated by the Bi $p_x$ and $p_y$ orbitals, so we ignore the $p_z$ orbital in the TB analysis.
Using the relevant $p_x$ and $p_y$ orbitals of each Bi atom as basis, the Bloch function is
\begin{eqnarray*}
\Phi_\mathbf{k}(\mathbf{r})=\frac{1}{\sqrt{N}} \sum_{\mathbf{R},\alpha=x,y}
e^{i\mathbf{k}\cdot{\mathbf{R}}} \left [ \phi^A_\alpha(\mathbf{r}-\mathbf{R})
+ e^{i\mathbf{k}\cdot{\mathbf{\tau}}} \phi^B_\alpha(\mathbf{r}-\mathbf{R-\tau}) \right ],
\end{eqnarray*}
where $\mathbf{\tau}$ is given by ($a$/2,$a$/2,0). 
$\phi_x^i(\mathbf{r})$ and $\phi_y^i(\mathbf{r})$ are
the atomic $p_x$ and $p_y$ orbitals at the Bi$^i$ atom with $i=A$ and $B$.
We use an orthogonal basis set and consider the nearest neighbor interaction only.
The resulting TB Hamiltonian for each $\mathbf{k}$
has following form of dimension 4.
\begin{equation*}
H_0=
\left ( \begin{array}{cccc}
 \epsilon_p & 0          & V_{xx}  & V_{xy}   \\
 0          & \epsilon_p & V_{yx}  & V_{yy}   \\
 V_{xx}  & V_{xy}  & \epsilon_p & 0           \\
 V_{yx}  & V_{yy}  & 0          & \epsilon_p  \\
\end{array} \right )
\end{equation*}
Here $\epsilon_p$ is the onsite energy of the $p_x$ and $p_y$ orbitals.
The hopping term $V_{\alpha\beta}$ is
\begin{eqnarray*}
V_{\alpha\beta}=\sum_{<\tau>} e^{i\mathbf{k}\cdot{\mathbf{\tau}}} \left <\phi_\alpha^A(\mathbf{r}) | H_0
| \phi_\beta^B(\mathbf{r}-\tau) \right >,
\end{eqnarray*}
where $\left <\tau \right >$ means a summation over the four nearest neighbors.
Employing the Slater--Koster parametrization,\cite{SK}
we obtain $V_{\alpha\beta}$:
\begin{eqnarray*}
V_{xx}=V_{yy}= 2 (t_{1\pi}+t_{1\sigma}) \cos (k_x a/2) \cos (k_y a/2) \\
V_{xy}=V_{yx}= 2 (t_{1\pi}-t_{1\sigma}) \sin (k_x a/2) \sin (k_y a/2).
\end{eqnarray*}
Here, $t_{1\pi}$ and $t_{1\sigma}$ are the hopping parameters for the $\pi$ and $\sigma$ bonds.

One can express $H_0$ in a simpler form by using the 2$\times$2 identity matrix $I$ and the Pauli matrix
$\sigma_x$.
\begin{equation*}
H_0=
\left ( \begin{array}{cc}
 \epsilon_p I & V_{xx} I +  V_{xy} \sigma_x  \\
 V_{xx} I +  V_{xy} \sigma_x & \epsilon_p I  \\
\end{array} \right )
\end{equation*}
The orbitals in the two Bi types interact only through the off-diagonal blocks $V_{xx} I +  V_{xy} \sigma_x$
whose eigenvalues are $V_{xx} + \gamma^{orb} V_{xy}$ with the relative phase $\gamma^{orb}=\pm 1$ 
between the $p_x$ and $p_y$ orbitals.
Thus the eigenvalues of $H_0$ can be written as  $\epsilon_p+\gamma^{atom} (V_{xx} + \gamma^{orb} V_{xy})$,
where $\gamma^{atom}=\pm 1$ denotes the relative phase between the two Bi types.

\begin{figure}[t]
\begin{center}
\includegraphics[width=8.5cm,angle=0]{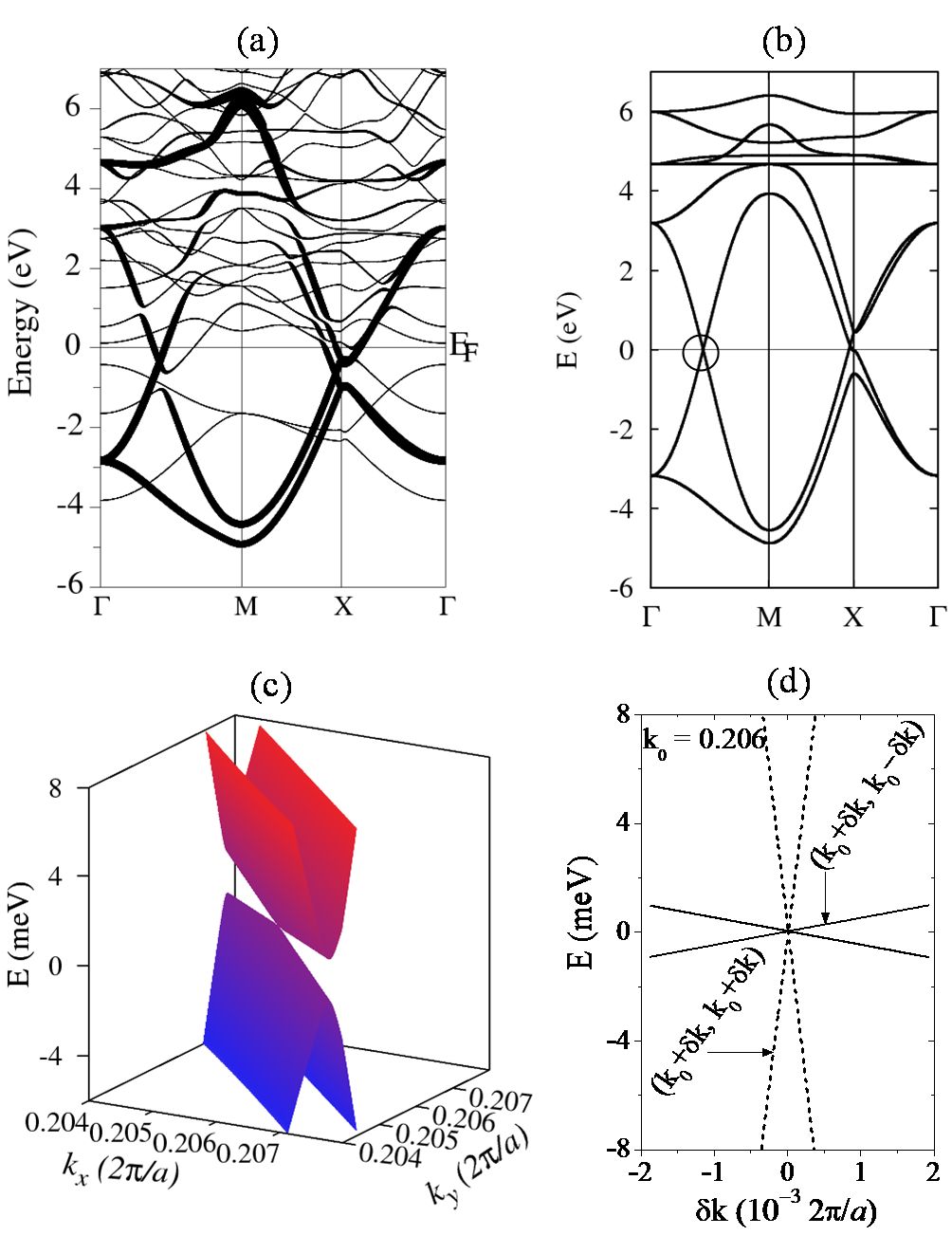}
\caption{ (color online) 
The band structures of the SrBi lattice shown in Fig. 5(a).
(a) The DFT band structure showing the contributions of the Bi 6$p_x$ and 6$p_y$ orbitals.
(b) The TB band structure. (c) $E(k_x,k_y)$ plot near the Dirac point.
(d) The anisotropic Dirac dispersion showing the bands along the $\Gamma$-M direction $(k_0+\delta k,k_0+\delta k)$ 
and its perpendicular direction $(k_0+\delta k,k_0-\delta k)$.}
\end{center}
\end{figure}

We estimated the values of $t_{1\sigma}$ {and} $t_{1\pi}$ 
as having a good agreement with the DFT band structures. 
The DFT calculation of the Bi square net was performed by considering well separated layers of the Bi square net, and the result is shown in Fig. 6(a).
Bi $p_x$ and $p_y$ driven DFT band structures are well reproduced by using $t_{1\sigma}=2.0$ eV, $t_{1\pi}=-0.5$ eV.
The TB band structure with $\epsilon_p=0$ is shown in Fig. 6(b). 
Additional DFT bands from the $p_z$ obital crossing the Fermi level are
not considered in this TB analysis.

The linear crossing at the Fermi level is found in Fig. 6(b) due to a folding of the $p_x$ and $p_y$ bands
from two Bi atoms.
Such degeneracy appears at every $\mathbf{k}$ point in the Fermi level
and produces a line-shape FS shown in Fig. 6(c).
So the unit-cell doubling in the Bi square net makes the conduction and valence bands touch each other 
on the whole Fermi surface.

\begin{figure}[t]
\begin{center}
\includegraphics[width=8.5cm,angle=0]{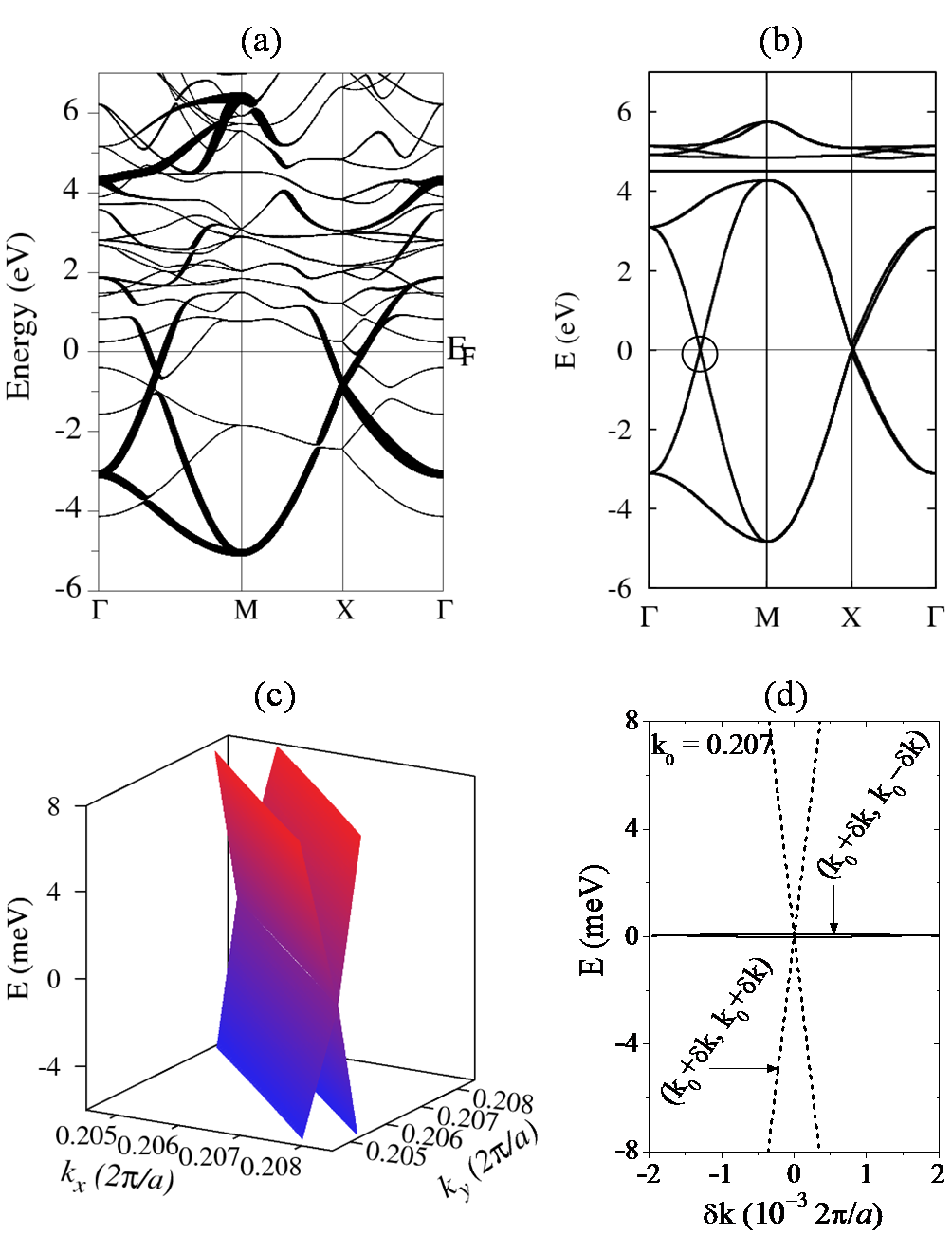}
\caption{ (color online) 
The band structures of the CaBi lattice shown in Fig. 5(b).
The scheme is same as that in Fig. 7, so refer to the caption of Fig. 7.}
\end{center}
\end{figure}

We now consider Sr atoms located at $ (0,a/2,\pm c)$ in the unit cell, as shown in Fig. 5(a).
According to the DFT results of SrMnBi$_2$, it is mainly the Sr $d$ orbitals which participate in the hybridization with the Bi $p$ band near the Fermi level.
The major contribution  of the Sr $d$ orbitals comes from the ${z^2}$ ($\sim$20\%) 
and ${xz/yz}$ ($\sim$50\%) orbitals, but
there are also non-negligible contributions from ${x^2-y^2}$ and ${xy}$ ($\sim$30\%).
So we consider all five $d$ obritals for the TB Hamiltonian.

The Hamiltonian matrix of the SrBi lattice, whose dimemsion is 14, is given below involving $H_0$.
\begin{equation*}
H_{SrBi}=
\left ( \begin{array}{cc}
 H_0 & V_{SrBi} \\
 V_{SrBi}^\dag & \epsilon_d I
\end{array} \right )
\end{equation*}
Here, $\epsilon_d$ denotes the onsite energy of the Sr $d$ orbital, and $I$ is the 10$\times$10 identity matrix.
$V_{SrBi}$, the hopping term between the Sr $d$ and Bi $p$ orbitals, is the 4$\times$10 matrix listed in Table III.

Fig. 7(a) shows the DFT band structure of the SrBi layer, 
again adopting periodic boundary condition along the $z$-direction with enough of a vacuum region. 
We choose $\epsilon_d = 4$ eV to represent the Sr $d$ states in the TB Hamiltonian,
and $u_\sigma = 1.5$ eV, $u_\pi = -0.5$ eV in $V_{SrBi}$. As shown in Fig. 7(b), 
a qualitative agreement between the DFT and the TB results is obtained. 
Around the linear crossing point, denoted by the circle in Fig. 7(b), the energy surfaces of two bands are plotted in Fig. 7(c).
One can see that the degeneracy along the band crossing line is lifted except one point 
which induces the anisotropic Dirac cone, as is clearly shown in Fig. 7(c).
The anisotropy of the momentum-dependent Fermi velocity is estimated to within an order of 10, as shown in Fig. 7(d).

\begin{table*}
\caption{Matrix elements of $V_{SrBi}$ with coefficients $c_1=(\sqrt{3}/\sqrt{2})u_{\sigma}$,
$c_2=\sqrt{2}u_{\pi}$, $c_3=(\sqrt{2}/4)u_{\sigma}-(\sqrt{3}/\sqrt{2})u_{\pi}$,
$c_4=(\sqrt{6}/4)u_{\sigma}+(1/\sqrt{2})u_{\pi}$, where $\kappa_x=k_xa/2$ and $\kappa_y=k_ya/2$.
We have used $u_{\sigma}=1.5$, $u_{\pi}=-0.5$ for the $\sigma$, $\pi$ bonds, respectively,
between Sr $d$ and Bi $p$ orbitals, where the resulting band structure is given in Fig. 7.}
\begin{tabular}{cc|ccccc|ccccc} \hline
&  & Sr$^A$ $(0,\frac{a}{2},\frac{a}{2})$ & & & & & Sr$^B$ $(0,\frac{a}{2},-\frac{a}{2})$ & & & & \\
&  & $xy$ & $yz$ & $xz$ & $x^2-y^2$ & $z^2$ & 
     $xy$ & $yz$ & $xz$ & $x^2-y^2$ & $z^2$ \\ \hline
Bi$^A$ $(0,0,0)$ & $x$ & 
0 & 0 & $c_1 \cos\kappa_x$ & $c_4 i\sin\kappa_x$ & $c_3 i\sin\kappa_x$ & 
0 & 0 & $-c_1 \cos\kappa_x$ & $c_4 i\sin\kappa_x$  & $c_3 i\sin\kappa_x$ \\ 
& $y$ &
$c_2 i\sin\kappa_x$ & $c_2 \cos\kappa_x$ & 0 & 0 & 0 &  
$c_2 i\sin\kappa_x$ & $-c_2 \cos\kappa_x$ & 0 & 0 & 0 \\ 
\hline
Bi$^B$ $(\frac{a}{2},\frac{a}{2},0)$ & $x$ & 
$c_2 i\sin\kappa_y$ & 0 & $c_2 \cos\kappa_y$ & 0 & 0 &  
$c_2 i\sin\kappa_y$ & 0 & $-c_2 \cos\kappa_y$ & 0 & 0 \\ 
& $y$ &
0 & $c_1 \cos\kappa_y$ & 0 & $-c_4 i\sin\kappa_y$ & $c_3 i\sin\kappa_y$ & 
0 & $-c_1 \cos\kappa_y$ & 0 & $-c_4 i\sin\kappa_y$  & $c_3 i\sin\kappa_y$ \\ \hline
\end{tabular}
\end{table*}

\begin{table*}
\caption{Matrix elements of $V_{CaBi}$ with coefficients $c_1=(\sqrt{3}/\sqrt{2})u_{\sigma}$,
$c_2=\sqrt{2}u_{\pi}$, $c_3=(\sqrt{2}/4)u_{\sigma}-(\sqrt{3}/\sqrt{2})u_{\pi}$,
$c_4=(\sqrt{6}/4)u_{\sigma}+(1/\sqrt{2})u_{\pi}$, where $\kappa_x=k_xa/2$ and $\kappa_y=k_ya/2$.
We have used $u_{\sigma}=1.2$, $u_{\pi}=-0.5$ for the $\sigma$, $\pi$ bonds, respectively,
between Ca $d$ and Bi $p$ orbitals, where the resulting band structure is given in Fig. 8.}
\begin{tabular}{cc|ccccc|ccccc} \hline
&  & Ca$^A$ $(\frac{a}{2},0,\frac{a}{2})$ & & & & & Ca$^B$ $(0,\frac{a}{2},-\frac{a}{2})$ & & & & \\
&  & $xy$ & $yz$ & $xz$ & $x^2-y^2$ & $z^2$ & 
     $xy$ & $yz$ & $xz$ & $x^2-y^2$ & $z^2$ \\ \hline
Bi$^A$ $(0,0,0)$ & $x$ & 
0 & 0 & $-c_1 \cos\kappa_x$ & $c_4 i\sin\kappa_x$ & $c_3 i\sin\kappa_x$ &
$c_2 i\sin\kappa_y$ & 0 & $c_2 \cos\kappa_y$ & 0 & 0 \\ 
& $y$ &
$c_2 i\sin\kappa_x$ & $-c_2 \cos\kappa_x$ & 0 & 0 & 0 & 
0 & $c_1 \cos\kappa_y$ & 0 & $-c_4 i\sin\kappa_y$ & $c_3 i\sin\kappa_y$ \\ 
\hline
Bi$^B$ $(\frac{a}{2},\frac{a}{2},0)$ & $x$ & 
$c_2 i\sin\kappa_y$ & 0 & $-c_2 \cos\kappa_y$ & 0 & 0 & 
0 & 0 & $c_1 \cos\kappa_x$ & $c_4 i\sin\kappa_x$ & $c_3 i\sin\kappa_x$ \\ 
& $y$ &
0 & $-c_1 \cos\kappa_y$ & 0 & $-c_4 i\sin\kappa_y$ & $c_3 i\sin\kappa_y$ &
$c_2 i\sin\kappa_x$ & $c_2 \cos\kappa_x$ & 0 & 0 & 0 \\ \hline
\end{tabular}
\end{table*}

\begin{figure}[t]
\begin{center}
\includegraphics[width=8.5cm,angle=0]{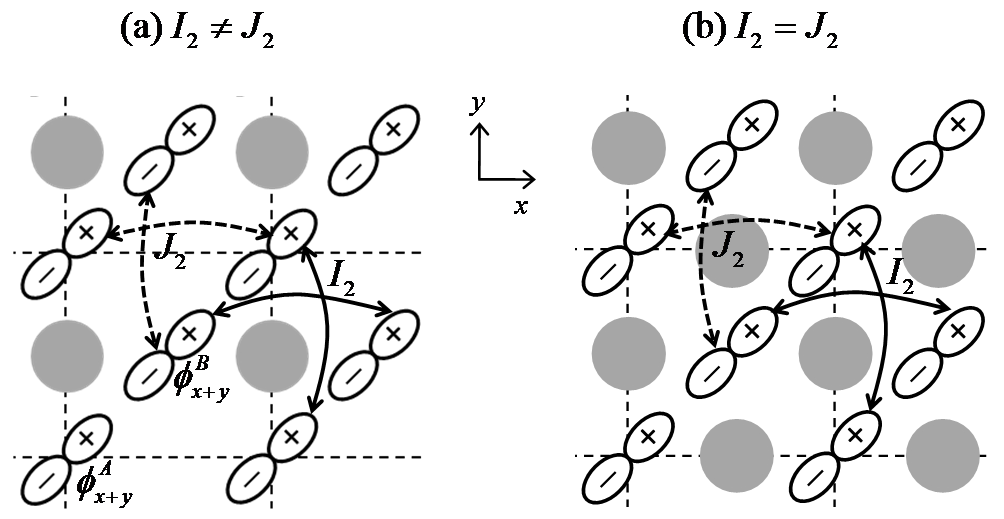}
\caption{ 
Illustration of local orbitals $\phi_{x+y}(\mathbf{r})$ on the Bi square net
and their next-nearest hopping strengths $I_2$ (solid) and $J_2$ (dashed)
which contribute to the formation of Dirac bands. 
The gray circles are the schematic representation of the perturbation potential 
by the stacked Sr or Ca layers.
(a) SrBi lattice has the overlapped stacking of Sr atoms near the Bi square net to induce different $I_2$ and $J_2$. (b) CaBi lattice has the alternate stacking of Ca atoms near the Bi square net to induce the same $I_2$ and $J_2$.}
\end{center}
\end{figure}

In the description of CaMnBi$_2$, the Ca atoms are located 
at $(a/2,0,c)$ and $(0,a/2,-c)$, as shown in Fig. 5(b).
Similarly, the Ca $d$ orbitals are hybridized with the Bi $p$ orbitals.
So a 14$\times$14 TB Hamiltonian $H_{CaBi}$ is given below with the onsite energy $\epsilon_d$ of the Ca $d$ orbital.
\begin{equation*}
H_{CaBi}=
\left ( \begin{array}{cc}
 H_0 & V_{CaBi} \\
 V_{CaBi}^\dag & \epsilon_d I
\end{array} \right )
\end{equation*}
The matrix elements of $V_{CaBi}$ are listed in Table IV.

In Figs. 8(a) and 8(b), we compare the band structures obtained by the DFT and TB methods, respectively.
Qualitative agreement is obtained by using the TB parameters $u_{\sigma}=1.2$ eV, $u_{\pi}=-0.5$ eV,
and $\epsilon_d = 4$ eV.
In contrast to the SrBi layer, the degeneracy along the line-type FS is not lifted, as can be  seen  from Fig. 8(c).
Fig. 8(d) clearly shows the almost zero gap along the continuous line $(k_0+\delta k,k_0-\delta k)$.

The distinct electronic structures between the SrBi and the CaBi lattices
arise from the different arrangements of the alkaline earth atoms with respect to the Bi square net.
This can be understood using perturbation theory for the first-order degenerate states. 
The two degenerate states along $(k_0+\delta k,k_0-\delta k)$ are unperturbed eigenstates.
The perturbation potential $V(\mathbf{r})$ by the adjacent $A$ (=Sr or Ca) atomic layers gives the following energy
eigenvalue difference (see the Appendix for a detailed derivation).
\begin{eqnarray*}
\delta \epsilon = 8(I_2-J_2) \sin (\delta k) \sim (I_2-J_2) \delta k, 
\end{eqnarray*}
where $I_2$ and $J_2$
are the coupling terms of $V$ between the next-nearest neighboring orbitals,
as illustrated in Fig. 9.
When two $A$ atomic layers have overlapped stacking as in the case of  SrBi, 
the perturbation $V$ results in $I_2 \ne J_2$ as shown in Fig. 9(a).
For this reason, a linear dispersion of $\delta\epsilon$ is obtained  
along $(k_0+\delta k,k_0-\delta k)$.  
This is consistent with the anisotropic Dirac cone feature of SrMnBi$_2$.
But, when two $A$ atomic layers have alternate stacking as in CaBi,
$I_2-J_2=0$, as is shown in Fig. 9(b). This causes $\delta \epsilon=0$.
So the degeneracy is not lifted, which is consistent with the result for CaMnBi$_2$.

\subsection{Chirality of Dirac electrons}

In the anisotropic Dirac band of the SrBi lattice,
the original two bands in the folded zone are associated with symmetric and antisymmetric 
combinations of two Bi sublattice orbitals.
In other words, the eigenstate is 
$(\phi_x^A+\gamma^{orb}\phi_y^A)+(\phi_x^B+\gamma^{orb}\phi_y^B)$ at a certain {\bf k}
on one side of the Dirac point, 
while it is $(\phi_{x}^A+\gamma^{orb}\phi_{y}^A)-(\phi_{x}^B+\gamma^{orb}\phi_{y}^B)$ on the other side.
A state with $\gamma^{atom}=1$ needs to be
changed continuously to another state with $\gamma^{atom}=-1$ around the Dirac point.
If we associate $\gamma^{atom}=\pm 1$ with up/down spinor states, 
the momentum {\bf k} and the (pseudo)spinor are coupled to each other, 
giving rise to a specific chirality.
Together with the linear variation of the energy eigenvalues with $\mathbf{k-k_0}$,
this coupling is generally described by the Weyl equation, which involves the Pauli matrices 
$\sigma_x, \sigma_y,\sigma_z$. 
Such a Hamiltonian gives a value for Berry's phase of $\pi$ when a state is scattered back to the original $\mathbf{k}$ 
state while going around $\mathbf{k_0}$.
For this reason, the back scattering is suppressed, as is known for graphene and materials with strong spin--orbit couplings.\cite{Ando}
This also can explain the abnormal phase observed in the quantum oscillation experiments.

In order to study the chirality,
we express an eigenstate as a superposition of $\phi_1=\phi_{x}^A+\gamma^{orb}\phi_{y}^A$, 
$\phi_2=\phi_{x}^B+\gamma^{orb}\phi_{y}^B$, that is, 
 $\psi_\mathbf{k}(\mathbf{r})=d_1 \phi_1(\mathbf{r}) +d_2\phi_2(\mathbf{r})=\left |d_1,d_2\right>$.
One needs to find 
a unitary matrix that transforms $\phi_1\pm\phi_2$ into spinor states with $\left <\sigma_z\right >=\pm1$. 
Such a transformation $T$ is given by
\begin{equation*}
T= \frac{1}{\sqrt{2}}
\left ( \begin{array}{rr}
 1 & -1  \\
 1 & 1  \\
\end{array} \right ).
\end{equation*}
For a given $\psi_\mathbf{k}(\mathbf{r})$, one obtains the expectation values of $\sigma_{\alpha}$ as
$\left <\sigma_\alpha \right >=\left <d_1,d_2\right | T^\dagger \sigma_\alpha T \left | d_1,d_2 \right >$ with
$\alpha=x,y,z$.
For $\alpha=x,z$, they are evaluated as 
$\left <\sigma_x \right >= d_1^\ast d_1 - d_2^\ast d_2$,
$\left <\sigma_z\right>= -d_1^\ast d_2 - d_1 d_2^\ast$.
For $\alpha=y$, $\left <\sigma_y\right>=i(d_1d_2^\ast -d_1^\ast d_2)$, which vanishes as long as $d_1$, $d_2$ have the same imanginary value.
Near the Dirac point, we calculate $n_x=\left <\sigma_x\right >, n_z=\left <\sigma_z\right >$
for each of the hole and electron TB states, to represent two-component vector fields
$(n_x,n_z)$. 
As shown in Fig. 10(a), there are four Dirac points in the first Brillouin zone, $(\pm k_0,\pm k_0)$
with $k_0>0$. 
Also we define local $k_1$ and $k_2$ axes for each Dirac point as indicated in Fig. 10(a).
As a function of $k_1$ and $k_2$, the vector field $(n_x,n_z)$ is represented in Figs. 10(b) and 10(c).
For all cases, 
the vector rotates by 2$\pi$ along the closed $\mathbf{k}$ loop, i.e., it has a winding number of 1.
Also when $k_2=0$, the hole is purely a $\left <\sigma_z\right >=1 (-1)$ state at negative (positive) $k_1$ 
as we expect, but it is reversed for the electron case.
So the chirality is the opposite for the hole state from the electron state.
From Fig. 10(b), the arrow rotates clockwise as one follows the contour line counter-clockwise
in both hole and electron states around ($k_0$, $k_0$) or ($-k_0$, $-k_0$).
In contrast, around ($-k_0$, $k_0$) or ($k_0$, $-k_0$), the rotation is the opposite, as shown in Fig. 10(c).

\begin{figure}
\begin{center}
\includegraphics[width=8.cm,angle=0]{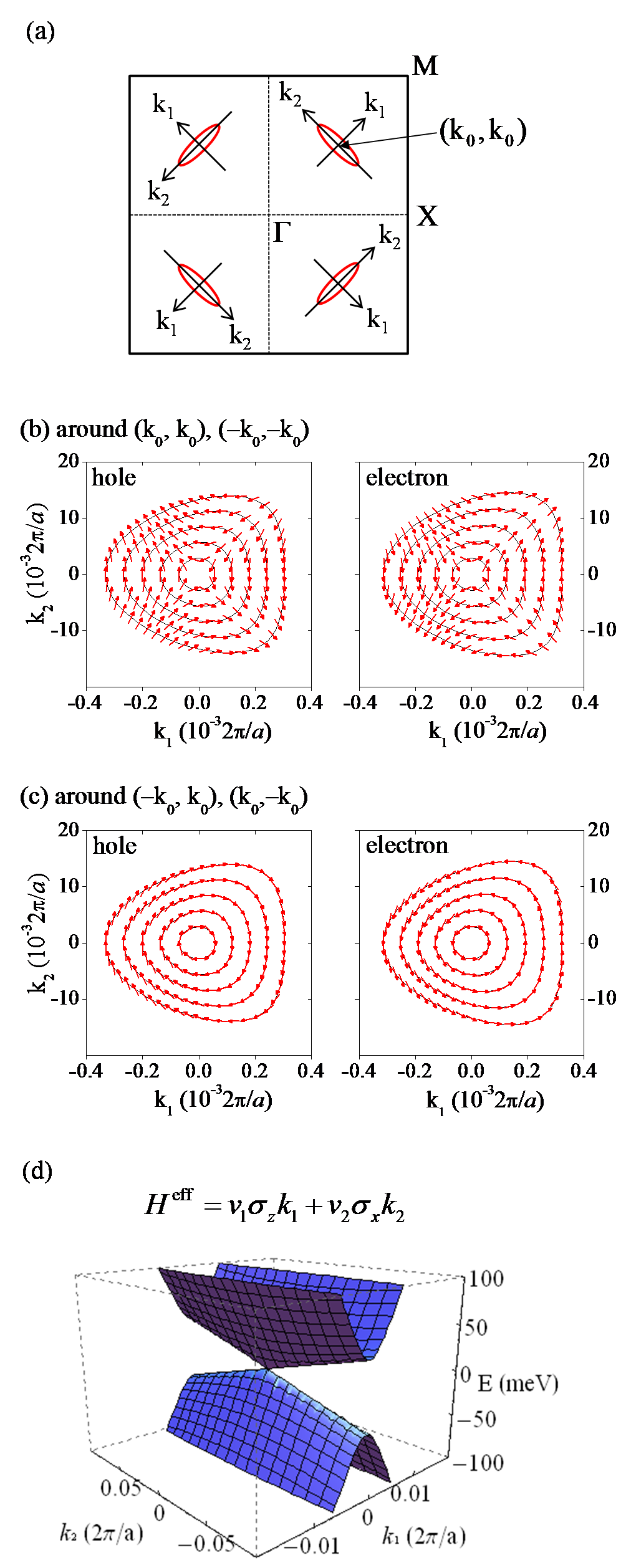}
\caption{(color online) 
(a) Definition of local $k_1$ and $k_2$ axes with the origin at each of four Dirac points 
in the first Brillouin zone. 
The Dirac point is at the center of the elliptical schematic energy contour, where
the first quadrant one is at $(k_0,k_0)$ with $k_0>0$.
(b) and (c) show $(n_x,n_z)=(\left <\sigma_x\right >, \left <\sigma_z\right >)$ characters 
of the hole and electron states of the SrBi lattice by the TB method 
near the Dirac points $\pm (k_0,k_0)$ and $\pm (-k_0,k_0)$, respectively.
The energy contours are built with an energy increase or decrease step of 0.001 eV relative to the Dirac point.
(d) hole and electron energy surfaces of the effective Hamiltonian assuming zero Dirac point energy.}
\end{center}
\end{figure}

One can construct the effective Hamiltonian 
$H^{eff}=H_\mathbf{k}-H_\mathbf{k_0}$ as a function of $\delta\mathbf{k}=\mathbf{k}-\mathbf{k_0}$
\begin{equation*}
H_\mathbf{k}-H_\mathbf{k_0}=\sum_{\alpha=0,x,y,z}\delta\mathbf{k}\cdot
\mathbf{v}_\alpha \sigma_\alpha,
\end{equation*}
where $\mathbf{v}_\alpha$ with $\alpha=x,y,z$ are the velocity parameters
and $\sigma_0$ is the unit matrix.\cite{PRL,JPSJ}
Nonvanishing $\mathbf{v}_0$ means the tilting of the Dirac cone, i.e., electron--hole asymmetry.
For the case of SrBi without an SOC, $\mathbf{v}_0=0$,
we simply use the $(n_x,n_z)$ results of Fig. 10 and assume an ideal elliptical shape for the energy contours in the limit 
$k_1,k_2 \rightarrow 0$
in order to obtain an approximate form of $H^{\rm eff}$ as below.
\begin{equation*}
H^{\rm eff}(k_1,k_2)=  v_1 \sigma_z k_1 +  v_2 \sigma_x k_2 = 
\left ( \begin{array}{rr}
 v_1 k_1 & v_2 k_2 \\
 v_2 k_2 & -v_1 k_1
\end{array} \right ),
\end{equation*}
where $v_1$ and $v_2$ are the Fermi velocities along the local unit vectors in the $k_1$ and $k_2$ directions, respectively.
As shown in Fig. 10(d),
positive (negative) eigenvalues of $H^{\rm eff}$ make up the upper (lower) part of the anisotropic Dirac 
cone. By fitting to the TB results, the absolute magnitudes of $v_1$ and $v_2$ are 
approximately 16 and 0.4 eV/(2$\pi/a$), respectively.
The sign of $v_1$ is always positive, but $v_2>0$ near $\pm (k_0,k_0)$ and $v_2<0$ near $\pm (-k_0,k_0)$.

\subsection{Spin--orbit coupling}

\begin{figure}[t]
\begin{center}
\includegraphics[width=8.5cm,angle=0]{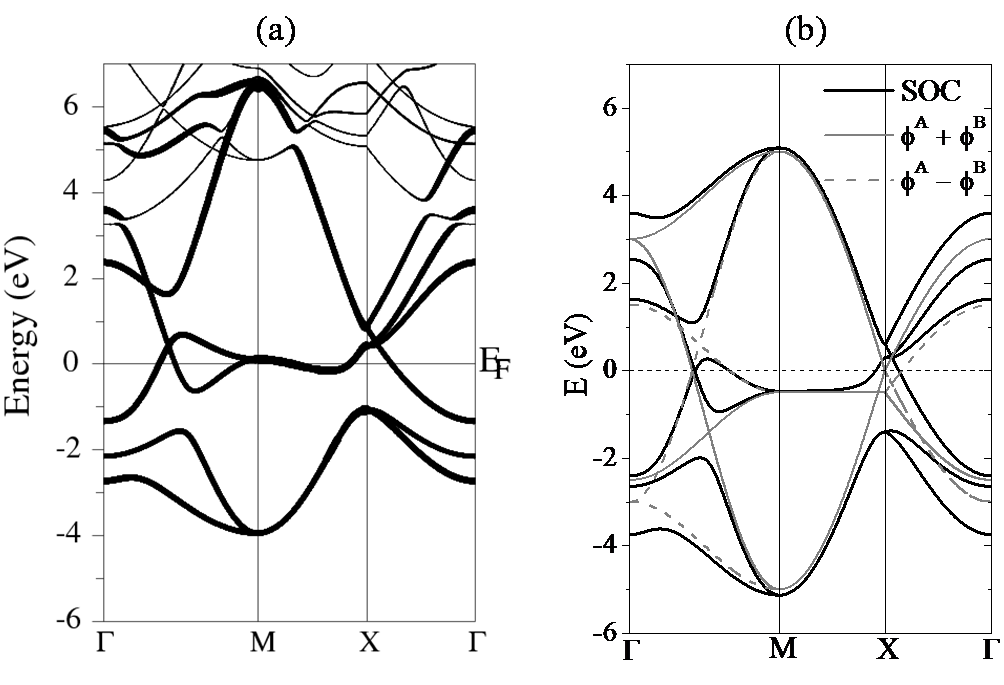}
\caption{ 
The band structures of the single layer Bi square net with SOC by (a) the DFT method and (b) the TB method. 
In (b) the sublattice symmetry $\gamma_{atom}=+1$ and $-1$ is indicated by the gray solid and dashed lines,
respectively, when the SOC is not considered.}
\end{center}
\end{figure}

The spin--orbit interaction is typically described by the following potential.
\begin{equation*}
H_{SO}(\mathbf{r})=\frac{1}{m_e^2 c^2} \frac{1}{r} \frac{dU(r)}{dr} \mathbf{L}\cdot\mathbf{S}
\end{equation*}
Replacing the radial integration by an effective constant $\lambda_{SO}$, 
we assume $H_{SO}=\lambda_{SO}\mathbf{L}\cdot\mathbf{S}$. 

The TB band structure of the single layer Bi square net is calculated with an additional term $H_{SO}$.
The matrix elements of $\mathbf{L}\cdot\mathbf{S}$ need to be found in the $\phi_{x,y,z}$ basis,
instead of usual spherical harmonics representation, $Y_l^m$.
By using the relationships 
$\phi_x=(Y_1^{1}+Y_1^{-1})/\sqrt{2}$, 
$\phi_y=i(Y_1^{1}-Y_1^{-1})/\sqrt{2}$,
$\phi_z=Y_1^{0}$, we can express $H_{SO}$ in terms of the following basis for each of Bi$^i$ ($i$ = $A$ and $B$),
\begin{equation*}
\left |\phi^{i}_{x,\uparrow} \right >, 
\left |\phi^{i}_{y,\uparrow} \right >, 
\left |\phi^{i}_{z,\uparrow} \right >,
\left |\phi^{i}_{x,\downarrow} \right >, 
\left |\phi^{i}_{y,\downarrow} \right >, 
\left |\phi^{i}_{z,\downarrow} \right >,
\end{equation*}
where $\uparrow (\downarrow)$ indicates the spin.
With this basis, we obtain $6\times 6$ $H_{SO}$ for each atom.
\[
H_{SO} = \lambda_{SO}
\left ( 
\begin{array}{rrrrrr}
      0 & i & 0 & 0 & 0 & 1 \\
     -i & 0 & 0 & 0 & 0 & i \\
      0 & 0 & 0 &-1 &-i & 0 \\
      0 & 0 &-1 & 0 &-i & 0 \\
      0 & 0 & i & i & 0 & 0 \\
      1 &-i & 0 & 0 & 0 & 0 
\end{array} \right )
\]


$H_{SO}$ is added 
to the $12\times 12$ $H_0$ which additionally takes into account
the $p_z$ orbital and the spin degree of freedom.
The calculatd TB band structure is shown by the solid line in Fig. 11(b),
where $\lambda_{SO}=0.6$, $\epsilon_{x,y}=0.0$, $\epsilon_z=-0.5$.
Also we have used $t_{1z} \cos (k_xa/2) \cos(k_ya/2)$ with $t_{1z}=2t_{1\pi}=-1$ for the $\pi$ interaction 
between $\phi_z^A$ and $\phi_z^B$.
Excellent agreement with the DFT band structure is shown in Fig. 11(a).

The SOC-induced splitting does not need to occur at every $\mathbf{k}$ point where
the $p_{x,y}$ band crosses the $p_z$ band.
In order to explain this, we show the $\gamma_{atom}=-1 (+1)$ bands whose local orbitals are given
by $\phi_{x+y}^A+\gamma_{atom}\phi_{x+y}^B$ and $\phi_z^A+\gamma_{atom}\phi_z^B$ without taking into consideration $H_{SO}$. 
Those bands have no hybridization with each other, as can be seen in Fig. 11(b).
Along the direction from $\Gamma$ to M, each $\phi_z$ band crosses the $\phi_{x+y}$ bands twice.
The leading contribution of the perturbation $H_{SO}$ is of second order, which is given
by $H_{SO}^{(2)} = \sum_m^{\epsilon_m \ne \epsilon_n} \left | \left < \psi_m^{(0)}|H_{SO}|\psi_n^{(0)} \right > \right |^2 /\left (\epsilon_n^{(0)}-\epsilon_m^{(0)} \right )$.
The main factor is due to the fact that $\left < \psi_m^{(0)}|H_{SO}|\psi_n^{(0)} \right >$ is non-vanishing
only when the values of $\gamma_{atom}$ 
are equal to each other for $\psi_m^{(0)}$ and $\psi_n^{(0)}$.
For this reason, $H_{SO}^{(2)}$ breaks the degeneracy only at the point where two crossing bands 
have the same $\gamma_{atom}$.


\subsection{Discussion}

The linear crossing behavior along lines of high symmetry is quite common in (transition) metal compounds. But SrMnBi$_2$ possesses distinguishing features that cause the anisotropic Dirac fermions as observed in experiments. The main factors can be summarized as follows. First, the Bi $p$ band is folded due to the unit cell doubling. This gives rise to a linear crossing of folded bands without lifting the degeneracy, which is rather common. Second, the two-fold rotational symmetry of the perturbing potential by adjacent atomic layers lifts the degeneracy except at the Dirac point. Third, and most importantly, the anisotropic Dirac cone mainly contributes to the electronic properties at the Fermi level. We have seen that most of the other bands are absent near the Fermi level. This is because the Mn-related bands are well spin-polarized and separated away from the Fermi level due to the antiferromagnetic ordering. Also the Dirac cone exhibits an exceedingly high Fermi velocity compared to the other Fermi pockets, dominating the electron transport properties. In these respects, SrMnBi$_2$ is quite unique.

The chirality that we obtained for SrBi, i.e., in Fig. 10, is almost identical to that of graphene. A superficial difference of SrMnBi$_2$ from graphene is that the Dirac cone is highly anisotropic. So, the contribution of the anisotropy to the transport properties deserves further investigation. Also it is not clear how the two types of Dirac cone shown in Figs. 10(b) and 10(c) are related to each other. For example, two inequivalent Dirac cones in graphene are related by the time-reversal transformation. Furthermore, SrMnBi$_2$ contains a significant SOC of Bi, in contrast to graphene. As one can see in Fig. 3(a), it has a sizable energy gap at the Dirac point and a significant electron--hole asymmetry due to the SOC. In such a case, Berry’s phase cannot be quantized to be $\beta=\pi$, as it is in graphene. Rather it should show a deviation from $\pi$.

We have not included the interlayer interaction in the TB calculation of the SrBi and CaBi layers. Such an omission is reasonable from the highly two dimensional nature of the Fermi surfaces in the DFT result in Fig. 4. But it is far from being a completely two dimensional system, as one can see a little dump near the $k_z$=0 for the anisotropic Fermi pocket in Fig. 4. A slight dispersion will originate from the indirect coupling through the insulating MnBi layers. In spite of there being little interlayer interaction, the anisotropic Dirac behavior is equally reproduced by the DFT method, i.e., the DFT results shown in Figs. 3(a) and (b), in comparison with the TB results, Figs. 7(c) and 8(c), respectively. This suggests that the indirect coupling does not destroy the anisotropic Dirac nature, which explains the Dirac fermions observed in some three dimensional systems such as iron pnictides, topological insulators, organic conductors, and so on.

\section{Conclusion}

The DFT results show the presence of an anisotropic Dirac cone in SrMnBi$_2$.
But, in CaMnBi$_2$ the band crossing occurs along a continuous line in momentum space.
From the TB analysis, we conclude that
the anisotropic potential created by the Sr atoms is a main factor for the anisotropic Dirac band.
Our study indicates that the nature of the Dirac dispersion is sensitive to the 
nearest neighbor interaction in the Bi square net. So, it would be interesting to investigate the Dirac nature of
the Bi square net with further changes, such as structural distortions or magnetism.
 
\begin{acknowledgments}
This work was supported by the National Research Foundation of Korea (NRF)
funded by the Ministry of Education, Science and Technology
 (Grants Nos. 2011-0010186, 2010-0005669, 2012-013838, 2011-0030147, 2012-029709, R32-2008-000-10180-0).
\end{acknowledgments}

\section*{APPENDIX}

A single Bi square net layer is regarded as an unperturbed system.
At each $\mathbf{k}$ with $0\le k_x,k_y \le \pi/a$, two bands touch
at the Fermi wavevectors as shown in Fig. 6(c). Their symmetries are given by
$\gamma^{orb}=+1$ and $\gamma^{atom}=\pm 1$.
The unperturbed eigenvalues are $\epsilon^0_1(\mathbf{k})=V_{xx}+V_{xy}$ and 
$\epsilon^0_2(\mathbf{k})=-V_{xx}-V_{yy}$.
The only condition for the band crossing, i.e., $\epsilon^0_1(\mathbf{k})=\epsilon^0_2(\mathbf{k})$, is that $V_{xx}+V_{yy}=0$.
This results in $\tan(k_xa/2)\tan(k_ya/2)=(t_{1\sigma}+t_{1\pi})/(t_{1\sigma}-t_{1\pi})$.
Assuming $t_{1\pi}=0$ to simplify our problem, we obtain $\cos(k_xa/2+k_ya/2)=0$.
As a result, the $\mathbf{k}$ line of the band crossing satisfies $k_x+k_y=\pi/a$.
The Dirac point $\mathbf{k_0}=(k_0,k_0)$ has $k_0=\pi/2a$. 
Also, from $\gamma^{orb}=+1$, 
the unperturbed eigenstates can be written as a linear combination of local orbitals, such as
$\phi^A_{x+y}(\mathbf{R})=
(\phi_x(\mathbf{r}-\mathbf{R})+\phi_y(\mathbf{r}-\mathbf{R}))/\sqrt{2}$ and
$\phi^B_{x+y}(\mathbf{R})=
(\phi_x(\mathbf{r}-\mathbf{R}-\mathbf{\tau})+\phi_y(\mathbf{r}-\mathbf{R}-\mathbf{\tau}))/\sqrt{2}$
with the lattice vector $\mathbf{R}$ and $\mathbf{\tau}=(a/2,a/2,0)$.
So they are expressed by
\begin{equation*}
\psi^0_j(\mathbf{r})=\frac{1}{\sqrt{N}}\sum_{\mathbf R} \left [
 \phi^A_{x+y}(\mathbf{R})+\gamma^{atom}_j\phi^B_{x+y}(\mathbf{R}) e^{i\mathbf{k}\cdot\mathbf{\tau}}
\right 
] e^{i\mathbf{k}\cdot\mathbf{R}},
\end{equation*}
where $\gamma^{atom}_1=1$  and $\gamma^{atom}_2=-1$ give the bonding and anti-bonding states, respectively.

Now we consider the perturbation by the additional potential of $A$ (=Sr or Ca) atomic layers.
Each $A$ atomic potential is assumed to be an isotropic potential $v(r)$, to give the total potential
$V(\mathbf{r})=\sum_\mathbf{R} v(\mathbf{r}-\mathbf{R}-(a/2)\mathbf{\hat{y}})$ for the case
illustrated in Fig. 9.
In order to get the energy shift due to the perturbation, we need to evaluate
the matrix elements $V_{ij}$=$<\psi_i^0|V|\psi_j^0>$ for $i,j=1,2$.
\begin{eqnarray*}
V_{ij}=&&\frac{1}{N} \sum_{\mathbf{R_1},\mathbf{R_2}} e^{-i(\mathbf{R_1}-\mathbf{R_2})\cdot\mathbf{k}}\int d\mathbf{r} \\
&& \left [ \phi_{x+y}^{A}(\mathbf{R_1})+\gamma^{atom}_i \phi_{x+y}^B(\mathbf{R_1}) 
e^{-i\mathbf{k}\cdot\mathbf{\tau}}
\right ] \\
&& V(\mathbf{r}) \left [ \phi_{x+y}^A(\mathbf{R_2})+\gamma^{atom}_j\phi_{x+y}^B(\mathbf{R_2})
e^{i\mathbf{k}\cdot\mathbf{\tau}} \right ]
\end{eqnarray*}
By assuming a short-ranged function $\phi_{x+y}^{A,B}(\mathbf{r})$, we consider the overlap integrals 
up to the second nearest neighbors, which are defined as follows.

\noindent On-site:
\begin{eqnarray*}
I_0&=&\int d\mathbf{r} \phi_{x+y}^A(\mathbf{0})V(\mathbf{r})\phi_{x+y}^A(\mathbf{0}) \\
&=&\int d\mathbf{r} \phi_{x+y}^B(\mathbf{0})V(\mathbf{r})\phi_{x+y}^B(\mathbf{0})
\end{eqnarray*}

\noindent First nearest neighbor:
\begin{eqnarray*}
I_1&=&\int d\mathbf{r} \phi_{x+y}^A(\mathbf{0})V(\mathbf{r})\phi_{x+y}^B(\mathbf{0})\\
J_1&=&\int d\mathbf{r} \phi_{x+y}^A(\mathbf{0})V(\mathbf{r})
\phi_{x+y}^B(-a\mathbf{\hat{x}})\\
\end{eqnarray*}

\noindent Second nearest neighbor:
\begin{eqnarray*}
I_2&=&\int d\mathbf{r} \phi_{x+y}^A(\mathbf{0})V(\mathbf{r})\phi_{x+y}^A(a\mathbf{\hat{y}})\\
&=&\int d\mathbf{r} \phi_{x+y}^B(\mathbf{0})V(\mathbf{r})\phi_{x+y}^B(a\mathbf{\hat{x}})\\ 
J_2&=&\int d\mathbf{r} \phi_{x+y}^A(\mathbf{0})V(\mathbf{r})\phi_{x+y}^A(a\mathbf{\hat{x}})\\
&=&\int d\mathbf{r} \phi_{x+y}^B(\mathbf{0})V(\mathbf{r})\phi_{x+y}^B(a\mathbf{\hat{y}})
\end{eqnarray*}

Note that $I_2\neq J_2$ for the SrBi lattice, as illustrated in Fig. 9, 
but $I_2 = J_2$ for the CaBi case.
After a straightforward procedure, one can obtain the following matrix elements.
\begin{eqnarray*}
V_{11}&=&I_0+I_1\left[ 1+\cos(k_xa+k_ya)\right]\\
&&+J_1\left[ \cos(k_xa)+\cos(k_ya)\right]\\
&&+(I_2+J_2)\left[ \cos(k_xa)+\cos(k_ya)\right]\\
V_{22}&=&I_0-I_1\left[ 1+\cos(k_xa+k_ya)\right]\\
&&-J_1\left[ \cos(k_xa)+\cos(k_ya)\right]\\
&&+(I_2+J_2)\left[ \cos(k_xa)+\cos(k_ya)\right]\\
V_{12}&=&V_{21}^{\ast}=2(I_2-J_2) \left[\cos(k_ya)-\cos(k_xa)\right]
\end{eqnarray*}
Along the  band crossing line with respect to the Dirac point $\mathbf{k_0}$,
that is, $\mathbf{k}=\frac{1}{a}(\frac{\pi}{2}+\delta,\frac{\pi}{2}-\delta)$,
this becomes $V_{11}=V_{22}=I_0$ with the only nontrivial element $V_{12}=V_{21}$.
Diagonalizing $V$ gives the following energy eigenvalue shifts.
\begin{eqnarray*}
\epsilon_i -\epsilon_i^0 = I_0 + (-1)^i 2(I_2-J_2) \left[\cos(k_ya)-\cos(k_xa)\right], i=1,2 
\end{eqnarray*}
By using $k_x=\frac{1}{a}\left (\frac{\pi}{2}+\delta \right )$ and 
$k_y=\frac{1}{a}\left (\frac{\pi}{2}-\delta \right )$, with $\delta \ll 1$, we obtain the following relation.
\begin{eqnarray*}
\delta \epsilon= \epsilon_2 -\epsilon_1 = 8 (I_2-J_2) \sin\delta \sim 8 (I_2-J_2)\delta,
\end{eqnarray*}
which has been double-checked by a numerical calculation.

\end{document}